\documentclass[12pt,twoside]{article}
\usepackage{latexsym}
\usepackage{amssymb}
\title{Perfect electromagnetic conductor}
\author{September 2004\\ I.V. Lindell, A.H. Sihvola} 
\date{Electromagnetics Laboratory\\ Helsinki University of Technology\\ PO Box 3000 Espoo, Finland 02015HUT}
\parskip=\medskipamount
\parskip=\medskipamount
\def\e{\begin{equation}} 
\def\f{\end{equation}} 

\def\##1{{\bf #1\mit}}
\def\%#1{{\mbox{\boldmath $#1$}}}

\def\=#1{{\overline{\overline{\mathsf #1}}}}

\def\SE{{\mathbb E}}
\def\SF{{\mathbb F}}

\def\/{\over}
\def\*{^{\displaystyle*}}

\def\.{\cdot}
\def\x{\times}
\def\:{\over}
\def\oo{\infty}

\def\ra{\rightarrow}

\def\l#1{\label{eq:#1}}
\def\r#1{(\ref{eq:#1})}
\def\am{\left(\begin{array}{c}}
\def\amm{\left(\begin{array}{cc}}
\def\ammm{\left(\begin{array}{ccc}}
\def\ammmm{\left(\begin{array}{cccc}}
\def\a{\end{array}\right)}

\def\add{\left|\begin{array}{cc}}
\def\addd{\left|\begin{array}{ccc}}
\def\adddd{\left|\begin{array}{cccc}}
\def\ad{\end{array}\right|}

\def\A{\alpha}
\def\B{\beta}

\def\E{\epsilon}
\def\g{\gamma}

\def\h{\eta}

\def\M{\mu}

\def\t{\tau}
\def\z{\zeta}

\def\TH{\theta}

\def\VR{\varrho}

\def\bi{\bibitem}

\def\W{\wedge}

\def\¤#1{\underline{\bf #1\mit}}

\def\bi{\begin{itemize}}

\def\ei{\end{itemize}}

\begin{document}

\maketitle

\subsection*{Abstract}

In differential-form representation, the Maxwell equations are represented by simple differential relations between the electromagnetic two-forms and source three-forms while the electromagnetic medium is defined through a constitutive relation between the two-forms. The simplest of such relations expresses the electromagnetic two-forms as scalar multiples of one another. Because of its strange properties, the corresponding medium has been considered as nonphysical. In this study such a medium is interpreted in terms of the classical Gibbsian vectors as a bi-isotropic medium with infinite values for its four medium parameters. It is shown that the medium is a generalization of both PEC (perfect electric conductor) and PMC (perfect magnetic conductor) media, with similar properties. This is why the medium is labeled as PEMC (perfect electromagnetic conductor). Defining a certain class of duality transformations, PEMC medium can be transformed to PEC or PMC media. As an application, plane-wave reflection from a planar interface of air and PEMC medium is studied. It is shown that, in general, the reflected wave has a cross-polarized component, which is a manifestly nonreciprocal effect.

\newpage
\setcounter{page}{1}

\section{Introduction}

Differential-form calculus is a branch of mathematics based on the algebra of multivectors (elements of space $\SE_p$) and dual multivectors (elements of space $\SF_p$) over an $n$-dimensional space of vectors \cite{Flanders,Deschamps,Difform}\footnote{The notation applied in the present paper coincides with that of \cite{Difform}.}. Its application to electromagnetic theory instead of the classical Gibbsian vector analysis is suggested by the simplicity and elegance obtained in writing the basic Maxwell equations as 
\e \#d\W\%\Phi=\%\g_m,\l{Max1} \f
\e \#d\W\%\Psi=\%\g_e. \l{Max2}\f 
Here, $\#d$ is the four-dimensional differential operator and $\W$ the exterior product, while 
\e \%\Phi= \#B + \#E\W\#d\t,\ \ \ \%\Psi= \#D - \#H\W\#d\t \l{PP} \f 
represent the four-dimensional electromagnetic two-forms (elements of the dual bivector space $\SF_2$) in terms of three-dimensional two-forms $\#B,\#D$ and one-forms $\#E,\#H$. The electric and magnetic source three-forms $\%\g_e, \%\g_m\in\SF_3$ are combinations of three-dimensional charge three-forms $\%\VR_e, \%\VR_m$ and current two-forms $\#J_e,\#J_m$ as
\e \%\g_e = \%\VR_e - \#J_e\W\#d\t,\ \ \ \ \ \%\g_m = \%\VR_m - \#J_m\W\#d\t. \f
Here $\t$ stands for the normalized time $\t=ct$. Denoting by $|$ the scalar product between multivectors and dual multivectors of the same grade \cite{Deschamps,Difform}, the most general linear relation between the two electromagnetic two-forms has the form
\e \%\Psi = \=M|\%\Phi, \l{M}\f
where $\=M$ is the medium dyadic involving 36 scalar parameters \cite{Difform}. Inserting \r{PP} in \r{M} and separating the spatial and temporal components, the medium equations can be represented in terms of a set of three-dimensional dyadics $\=\A,\=\E{}',\=\M,\=\B$ by 
\e \#D = \=\A|\#B + \=\E{}'|\#E, \f
\e \#H = \=\M{}^{-1}|\#B + \=\B|\#E. \f
The dyadic $\=\E{}'$ is in general not the same as $\=\E$ in the alternative representation
\e \#D = \=\E|\#E + \=\xi|\#H, \l{D}\f
\e \#B = \=\z|\#E + \=\M|\#H, \l{B}\f
corresponding to the same four-dimensional medium dyadic $\=\M$. In the Gibbsian representation we can apply medium equations similar to \r{D}, \r{B},
\e \#D = \=\E\.\#E + \=\xi\.\#H, \l{D1}\f
\e \#B = \=\z\.\#E + \=\M\.\#H, \l{B1}\f
for the vector-valued fields $\#D,\#B,\#E,\#H\in\SE_1$, in terms of the medium dyadics $\=\E,\=\xi,\=\z,\=\M$. One must note that the field and medium quantities in \r{D}, \r{B} and \r{D1}, \r{B1} are denoted by the same symbols even if they are elements of different spaces, because they represent the same physical quantities.

\section{Simple isotropic medium}

Obviously, the simplest electromagnetic medium as defined by \r{M} is obtained when the $\=M$ dyadic is a scalar factor $M$ so that the relation becomes
\e \%\Psi = M\%\Phi. \l{iso1}\f
In terms of three-dimensional one- and two-forms or Gibbsian vectors the relation \r{iso1} has the form 
\e \#D = M\#B,\ \ \ \ \#H = -M\#E. \l{iso2}\f
From the viewpoint of four-dimensional differential forms \r{iso1} appears to define the only possible isotropic medium in the sense that it is invariant in all possible affine transformations \cite{Difform}. This means that the medium has no special spatial direction in the four-dimensional space and it appears the same for all observers moving with constant velocity. In contrast, it is known that change of motion of the observer changes the medium known as isotropic in the Gibbsian representation \r{D1}, \r{B1} to a more general bi-anisotropic medium. While being mathematically simple, the medium defined by \r{iso1} appears physically very strange, because for constant $M$ a contradiction arises in the Maxwell equations \r{Max1} and \r{Max2}, unless the sources satisfy the special relation $\%\g_e=M\%\g_m$. In this special source system the two Maxwell equations become the same and the solution is not unique. This is why such a medium has been labeled as nonphysical \cite{Hehl,427}. It appears that the problem cannot be posed in such a simple way but one should consider the process of creating the sources in the medium more carefully. On the other hand, everywhere outside the source region in such a medium the Poynting two-form $\#E\W\#H$ and energy-density three-form $\#E\W\#D+ \#H\W\#B$ vanish, which means that there cannot exist any energy or transmission of electromagnetic power in such a medium.

Since the simple isotropic medium defined by \r{iso1} arises so naturally in the differential-form formalism, let us consider its representation in the Gibbsian vector formalism more closely. It turns out that the medium can be represented as a bi-isotropic medium defined in terms of four scalar medium parameters $\E,\xi,\z,\M$ as
\e \#D = \E\#E + \xi\#H, \l{medD} \f
\e \#B = \z\#E + \M\#H. \l{medB}\f
Actually, \r{iso1} is a special case of such a medium and the four parameters can be expressed in terms of just two parameters $M$ and $q$ as
\e \E = Mq,\ \ \xi=q,\ \ \z=q,\ \ \M=q/M. \l{E}\f
In fact, \r{medD} and \r{medB} now become
\e \#D = q(M\#E + \#H),\ \ \ \ \#B = q(\#E + \#H/M). \f
This shows us that one of the equations, $\#D=M\#B$, is satisfied for any values of $q$ and $M$ while the other equation $\#H=-M\#E$ requires that the parameter $q$ must become infinite:
\e q\ra\oo. \f
From \r{E} we see that, this being the case, all four medium parameters actually become infinite. Their relations can be expressed as
\e \xi = \z = \sqrt{\M\E} = q,\ \ \ \ \sqrt{\E/\M}=M. \f
This kind of a medium can be characterized as a special Tellegen medium (nonchiral and nonreciprocal bi-isotropic medium) \cite{Chibi}.

\section{PEMC medium}

To have some insight in the medium defined by \r{E}, let us perform a duality transformation which is known to transform a set of fields and sources to another set and the medium to another one. In its most general form, the duality transformation can be defined as a linear relation between the electromagnetic fields, and represented in terms of four scalar parameters $A,B,C,D$ as \cite{JEMWA98}
\e \am \#E\\ \#H\a_d = \amm A & B\\ C & D\a \am \#E\\ \#H\a,\ \ \ \ AD-BC\not=0. \f
Requiring that the Maxwell equations be transformed to Maxwell equations for the dual fields, the other two field vectors must be transformed as
\e \am \#D\\ \#B\a_d = \amm D & -C\\ -B & A\a \am \#D\\ \#B\a \f
and the medium parameters as 
\e \am \E \\ \xi\\ \z \\ \M\a_d = {1\over AD-BC}\ammmm  D^2&-CD &-CD &C^2 \\ -BD&AD &BC &-AC \\ -BD&BC &AD &-AC \\ B^2&-AB &-AB &A^2 \a \am \E \\ \xi\\ \z \\ \M\a. \l{dualM}\f
It is easy to see that for $\xi=\z$ we also have $\xi_d=\z_d$, which means that a Tellegen medium is transformed to another Tellegen medium. Now let us study whether there exist possible transformations leading to vanishing magnetoelectric parameters $\xi_d=\z_d=0$ for a given medium \r{E}. This requirement leads to the condition
\e -BDM + AD+BC - AC/M=(A - BM)(D-C/M)=0 \f
for the transformation parameters. Thus, there are two possible transformations denoted by subscripts 1 and 2 and defined by the conditions
\e A_1=B_1M,\ \ \ \ D_2=C_2/M. \f
Inserted in \r{dualM}, the two duality transformations lead to the respective two sets of transformed medium parameters with vanishing $\xi_d$ and $\z_d$. Remarkably, it appears that, in each case, three of the four parameters are transformed to zero:
\e \am \E \\ \xi\\ \z \\ \M\a_{1d} = \am  q(MD_1-C_1)/B_1M \\ 0 \\ 0 \\ 0 \a, \f
\e \am \E \\ \xi\\ \z \\ \M\a_{2d} = \am  0 \\ 0 \\ 0 \\ q(A_2-B_2M)/D_2M \a. \f
From the assumption $AD-BC\not=0$ we must also have $MD_1-C_1\not=0$ and $A_2-B_2M\not=0$. Thus, for $q\ra\oo$ the nonzero transformed medium elements become infinite in both cases. 

This shows us that it is possible to find a duality transformation which transforms the bi-anisotropic medium defined by \r{E} and $q\ra\oo$ to either a PEC (perfect electric conductor) medium satisfying \cite{Stevenson}
\e \E_{1d}\ra\oo,\ \ \ \ \xi_{1d}=\z_{1d}=\M_{1d}=0,\f
or to a PMC (perfect magnetic conductor) medium satisfying
\e \E_{2d}=\xi_{2d}=\z_{2d}=0,\ \ \ \ \M_{2d}\ra\oo. \f
Actually, PEC and PMC are special cases of the medium \r{M}. In fact, because in the PMC medium we have $\#D=\E\#E+\xi\#H=0$ and $\#H = (\#B-\z\#E)/\M=0$ in the Gibbsian vector representation, these correspond to vanishing of the two-form $\%\Psi=0$. Thus, the PMC condition can be characterized by the special parameter value $M=0$ in \r{M}. Similarly, the PEC medium corresponds to $\%\Phi=0$ and $M=\oo$. This gives us reason to call the more general medium defined by \r{M} as the perfect electromagnetic conductor (PEMC). It is a one-parameter class of media. Because the Poynting vector and energy density are transformed as
\e \#E_d\x\#H_d = (AD-BC)\#E\x\#H, \f
\e \#E_d\.\#D_d +\#H_d\.\#B_d = (AD-BC)(\#E\.\#D+\#H\.\#B), \f
in the general duality transformation and since these quantities vanish for PEC and PMC media, they also vanish in the PEMC medium.

Because the duality transformation can be inverted, the medium \r{M} can also be defined as one obtained from a PEC or a PMC medium through a special duality transformation. The effect of the duality transformation can be made more transparent by the following special choice of transformation parameters:
\e A = D=\cos(\TH/2),\ \ \ B = {1\over M}\sin(\TH/2),\ \ \ C = -M\sin(\TH/2). \f
In this case the PEMC medium parameters are transformed to those of another PEMC medium through the $\TH$ parameter as
\e \E_d = qM(1+\sin\TH),\ \ \ \xi_d=\z_d = q\cos\TH,\ \ \ \M_d = {q\over M}(1-\sin\TH). \f
It is seen that $\TH$ varying in the interval $-\pi/2\cdots\pi/2$ reduces the PEMC medium to PMC and PEC at the end points and the original medium is obtained at $\TH=0$. For all values of $\TH$ the medium satisfies the condition $\M_d\E_d=\xi_d\z_d$.

\section{Reflection from a PEMC boundary}

Because the PEMC medium does not allow electromagnetic energy to enter, an interface of such a medium serves as an ideal boundary to the electromagnetic field. Let us consider the boundary of PEMC medium and air with unit normal vector $\#n$. Because tangential components of the $\#E$ and $\#H$ fields are continuous at any interface of two media, one of the boundary conditions for the medium in the air side is
\e \#n\x(\#H + M\#E)=0, \f
because a similar term vanishes in the PEMC-medium side. The other condition is based on the continuity of the normal component of the $\#D$ and $\#B$ fields which gives another boundary condition as
\e \#n\.(\#D-M\#B)=0. \f
The latter condition can also be written as
\e \#n\.(\E_o\#E - M\M_o\#H)=0. \f
Put together, these can be expressed as the vector condition
\e M\#E = -\#H+ (M^2\h_o^2+1)\#n\#n\.\#H ,\ \ \ \ \h_o=\sqrt{\M_o/\E_o}. \l{BC}\f 
As a check, for $\M\ra\oo$ we obtain the PEC conditions $\#n\x\#E=0$ and $\#n\.\#H=0$.

As an application, let us consider plane-wave reflection from a PEMC boundary plane at $z=0$. For simplicity, the incident and reflected plane waves in the region $z<0$ are assumed polarized parallel to the boundary. The total fields are
\e \#E(z) = \#E^ie^{-jk_oz}+ \#E^re^{jk_oz},\f
\e \h_o\#H(z) = \#u_z\x\#E^i e^{-jk_oz} -\#u_z\x\#E^r e^{jk_oz}. \f
Inserted in the boundary conditions \r{BC} at $z=0$ with $\#n=-\#u_z$ we have
\e M\h_o(\#E^i+\#E^r) = -\#u_z\x(\#E^i -\#E^r), \f
or
\e (\#u_z\x\=I -M\h_o\=I)\.\#E^r = (\#u_z\x\=I + M\h_o\=I)\.\#E^i, \f
with $\=I = \#u_x\#u_x+\#u_y\#u_y$. Multiplying this by the dyadic $(\#u_z\x\=I +M\h_o\=I)\.$, the reflected field is obtained as
\e \#E^r = -{1\over1+ M^2\h_o^2}[(-1+M^2\h_o^2)\#E^i + 2M\h_o\#u_z\x\#E^i]. \f
This means that, for a linearly polarized incident field (real $\#E^i$), the field reflected from such a boundary has both a co-polarized component (multiple of $\#E^i$) and a cross-polarized component (multiple of $\#u_z\x\#E^i$) in the general case. For the PMC and PEC special cases ($M=0$ and $M=\oo$, respectively,) the cross-polarized component vanishes. For the special PEMC case $M=1/\h_o$, we have
\e \#E^i = -\#u_z\x\#E^i, \f
which means that the reflected field appears totally cross-polarized. Thus, the boundary acts as a twist polarizer which is a nonreciprocal device (\cite{Chibi}, p.84).

\section{Conclusion}

In this study, a class of electromagnetic media, defined in the simplest possible manner as 'the isotropic medium' when using the four-dimensional differential-form formalism, was given an interpretation in terms of the corresponding Gibbsian formalism. It turned out that the class can be defined as that of certain bi-isotropic media whose all four scalar parameters have infinite values and no power propagation in the medium is possible. Since this kind of one-parameter class of media could be shown to be a generalization of both PEC and PMC media, it can be called the class of PEMC media. Since a PEMC medium acts as an ideal boundary, plane-wave reflection was considered from a planar boundary. In the general case, the reflected wave has both a co-polarized and a cross-polarized component.


\begin{thebibliography}{99}

\bibitem{Flanders} H. Flanders, {\it Differential Forms}, New York: Academic Press, 1963.

\bibitem{Deschamps} G.A. Deschamps, "Electromagnetics and differential forms," {\it Proc.\ IEEE}, vol.69, no.6, pp.676--696, June 1981.

\bibitem{Difform} I.V. Lindell, {\it Differential Forms in Electromagnetics}, New York: Wiley and IEEE Press, 2004. 

\bibitem{Hehl} F.W. Hehl, Yu.N. Obukhov, {\it Foundations for Classical Electrodynamics} Boston: Birkh\"auser 2003, p.246.

\bibitem{427} I.V. Lindell, "Affine transformations and bi-anisotropic media in differential-form approach," {\it J. Electromag.\ Waves and Appl.}, vol.18, no.9, pp.1259-1273, 2004. 

\bibitem{Chibi} I.V. Lindell, A.H. Sihvola, S.A. Tretyakov, A.J. Viitanen, {\it Electromagnetic Waves in Chiral and 
Bi-Isotropic Media}, Boston: Artech House, 1994. 

\bibitem{JEMWA98} I.V. Lindell, L.H. Ruotanen, "Duality transformations and Green dyadics for bi-anisotropic media," {\it J. Electromag.\ Waves and Appl.}, vol.12, pp.1131-1152, 1998.
 
\bibitem{Stevenson} A.F. Stevenson, "Solution of electromagnetic scattering problems as power series in the ratio (dimension of scatterer)/wavelength," {\it J. Appl.\ Phys.}, vol.24, no.9, pp.1134-1142, 1953.


\end{thebibliography}
\end{document}